\documentclass[conference]{IEEEtran}
\IEEEoverridecommandlockouts
\usepackage{amsmath,amssymb,amsfonts}

\usepackage{cite}
\usepackage{amsmath,amssymb,amsfonts}
\usepackage{algorithm}
\usepackage{algorithmic}
\usepackage{graphicx}
\usepackage{booktabs}
\usepackage{multirow,enumitem}
\usepackage{textcomp}
\usepackage{xcolor}
\usepackage{bbm}
\usepackage{dsfont}
\usepackage[hidelinks]{hyperref}
\usepackage{tikz}
\usepackage{pgfplots}
\pgfplotsset{compat=1.18}
\usetikzlibrary{shapes.geometric,arrows.meta,positioning,fit,
                backgrounds,calc,decorations.pathreplacing}

\definecolor{clrC1}{RGB}{52,114,53}
\definecolor{clrC2}{RGB}{192,48,40}
\definecolor{clrC3}{RGB}{31,100,170}
\definecolor{clrOff}{RGB}{232,244,232}
\definecolor{clrOn} {RGB}{232,238,250}
\definecolor{clrAtk}{RGB}{252,235,233}
\definecolor{clrKPI} {RGB}{245,248,255}
\definecolor{clrGray}{RGB}{230,230,235}
\definecolor{clrT}    {RGB}{173,214,241}
\definecolor{clrClean}{RGB}{ 30,130, 60}
\definecolor{clrTrig} {RGB}{192, 48, 40}
\definecolor{clrE}    {RGB}{ 31,100,170}
\definecolor{clrEp}   {RGB}{140, 60,180}

\def\BibTeX{{\rm B\kern-.05em{\sc i\kern-.025em b}\kern-.08em
    T\kern-.1667em\lower.7ex\hbox{E}\kern-.125emX}}
 
\newcommand{\Eperp}{E^{\perp}}
\newcommand{\En}{E_n}
\newcommand{\proj}[1]{\mathrm{Proj}_{#1}}
\newcommand{\Real}{\mathbb{R}}
\newcommand{\Exp}{\mathbb{E}}
\newcommand{\pid}{\pi^{\dagger}}
\newcommand{\sd}{s^{\dagger}}
\newcommand{\norm}[1]{\left\lVert#1\right\rVert}

\newcommand{\basepanel}

\renewcommand{\baselinestretch}{0.87}

\begin{document}

\title{ORAN-DEFEND: Subspace Detection and Sanitization of Backdoor DRL xApps in Open RAN \thanks{This work is supported by National Science Foundation under Grant Numbers CNS-2202972, CNS- 2318726, and CNS-2232048.}}

\author{\IEEEauthorblockN{Md Raihan Uddin, Fatemeh Lotfi, Tolunay Seyfi, Fatemeh Afghah}
\IEEEauthorblockA{\textit{Department of Electrical and Computer Engineering, Clemson University, Clemson, SC, USA} \\
\{uddin2,flotfi,tseyfi,fafghah\}@clemson.edu}
}

\maketitle

\begin{abstract}

Open Radio Access Networks (O-RAN) increasingly delegate near-real-time control to deep reinforcement learning (DRL) xApps obtained from third-party vendors, creating a new supply-chain attack surface. A backdoor policy behaves optimally until an adversary injects a covert trigger into the observed key performance indicator (KPI) telemetry, at which point it issues harmful control actions that degrade quality of service (QoS). We present \textbf{ORAN-DEFEND}, a retraining-free wrapper that sanitizes a frozen, potentially compromised xApp by projecting each KPI window onto a safe subspace estimated from a small number of trusted clean rollouts via singular value decomposition (SVD). We establish, both analytically and empirically, a precise recovery condition: the defense succeeds if the trigger energy concentrates in the orthogonal complement of the safe subspace, and we quantify this boundary through the trigger's $\Eperp$ energy fraction. On the Colosseum COLORAN dataset, we evaluate four structurally distinct DRL backdoor attacks, like TrojDRL, SleeperNets, BadRL, and Q-Incept, spanning inner-loop and outer-loop poisoning regimes and demonstrate $100\%$ return recovery and $\geq99.5\%$ defense success rate across all four when the subspace assumption holds. A geometry ablation reveals an intrinsic and previously uncharacterized limit of any linear projection defense: when the trigger collocates with the legitimate signal, the $\Eperp$ energy fraction governs recovery monotonically, and the linear residual detector collapses to chance even while a nonlinear classifier retains perfect separability.

\end{abstract}
 
\begin{IEEEkeywords}
O-RAN, reinforcement learning security, backdoor attacks, subspace sanitization, trustworthy AI, xApp.
\end{IEEEkeywords}

\section{Introduction}
\label{sec:intro}

The Open Radio Access Network (O-RAN) initiative disaggregates the cellular radio access network into vendor-neutral, software-defined components coordinated by RAN Intelligent Controllers (RICs)~\cite{polese2023understanding}. The Near Real-Time RIC (Near-RT RIC) executes \emph{xApps closed-loop} control applications that consume telemetry from the key performance indicator (KPI) of the user equipment (UE) and make decisions regarding the scheduling, beam-management and allocation of radio resources on timescales $10\,\text{ms}$ -- $1\,\text{s}$. Deep reinforcement learning (DRL) has emerged as the dominant framework for synthesizing these xApp policies, owing to its capacity to optimize long-horizon quality-of-service (QoS) objectives directly from telemetry without an explicit network model~\cite{lacava2025poison,DORA_25,REAL25,Lotfi_WCNC25,MORPH}. The computational cost of training high-performing DRL agents, however, substantially exceeds what individual operators can sustain, driving a growing market in pretrained xApp policies distributed by third-party vendors and open model repositories. This procurement model introduces a qualitatively new attack surface.

\subsection{Related Work}
\textbf{Backdoor attacks in Machine Learning.}
Backdoor attacks, first demonstrated by BadNets~\cite{gu2019badnets}, insert a hidden trigger-label association during training so that a compromised model behaves normally on clean inputs but maps trigger-stamped inputs to an attacker-chosen output. Trojaning attacks~\cite{liu2018trojaning} extend this to internal neuron manipulation. The threat is acute whenever users obtain pretrained models from untrusted sources.

\textbf{Backdoor Attacks in Deep Reinforcement Learning.}
Extending backdoors to DRL is non-trivial: the adversary must corrupt the agent's \emph{policy} rather than a static classifier, and harm manifests through long-horizon cumulative return rather than a single misclassification. In what follows, $\pid$ denotes the resulting \emph{backdoor policy}. TrojDRL~\cite{kiourti2020trojdrl} establishes the foundational threat model via reward poisoning. SleeperNets \cite{rathbun2024sleepernets} improves stealthiness through provably universal value function manipulation. BadRL~\cite{cui2024badrl} minimizes poisoned steps via Q-function-sensitivity-based trigger selection. Q-Incept~\cite{rathbun2024adversarial} decouples training-time corruption from runtime delivery by injecting synthetic transitions directly into the replay buffer. Collectively, these works confirm that backdoor policies are a credible, multi-mechanism threat, but none addresses detection or mitigation.

\textbf{Backdoor Defenses}
Supervised-learning defenses (Neural Cleanse~\cite{wang2019neural}, Fine-Pruning~\cite{liu2018fine}, spectral signatures~\cite{gu2019badnets}) universally require white-box access or training data, neither of which is available when the xApp is a frozen proprietary binary \cite{owfi2025adaptattackdomainshift}. The closest prior work is Bharti \textit{et~al.}~\cite{bharti2022provable}, which proves a probably approximately correct (PAC)-style optimality guarantee for subspace-projection sanitization under the assumption that triggers reside in $\Eperp$ (the orthogonal complement $\Eperp$ of the safe subspace $E$). Their evaluation covers two Atari environments and a single attack family and does not characterize the boundary when the assumption is violated. ORAN-DEFEND extends this framework to O-RAN with four attack families and provides the first characterization of the defense boundary via the $\Eperp$ energy fraction.

\textbf{Security in O-RAN}
Polese \textit{et~al.}~\cite{polese2023understanding} survey O-RAN attack surfaces, identifying the xApp interface as high-risk. Lacava \textit{et~al.}~\cite{lacava2025poison} demonstrate practical backdoor feasibility on a real O-RAN testbed but focus on the attack side. KPI-injection attacks are studied in~\cite{alimohammadi2024kpi}; Long short-term memory (LSTM)-based anomaly detection in~\cite{moore2025anomaly}; and dynamic machine Learning-defense mechanisms in~\cite{kakani2025mitigating}. None provides a retraining-free, provably grounded defense for the backdoor-policy threat on O-RAN telemetry.

\subsection{Motivation and Contributions}
A malicious vendor can publish a \emph{backdoor policy}: a DRL agent indistinguishable from a benign expert under clean operation but harboring a covert trigger response association. At deployment, the adversary perturbs the live KPI stream, causing the policy to silently select harmful actions, degrading throughput, violating ultra-reliable and low-latency communications (URLLC) budgets, or inducing interference. While passive monitoring remains blind. Because the policy optimizes cumulative discounted return, a single triggered step can cascade into persistent QoS degradation far exceeding any instantaneous metric. Conventional defenses presuppose white-box access or retraining capability, neither available when the xApp arrives as a frozen proprietary binary. A viable defense must therefore operate as a \emph{telemetry-level wrapper} using only black-box inference and trusted clean rollouts.

We present \textbf{ORAN-DEFEND}, a provably grounded, retraining-free wrapper extending the subspace sanitization framework~\cite{bharti2022provable} to O-RAN. Nominal xApp operation concentrates the discounted state-occupancy in a low-dimensional \emph{safe subspace} $\En$ of the KPI feature space, estimable via SVD from clean rollouts. When the adversarial trigger resides in the orthogonal complement $\Eperp$ (subspace of $\Real^D$ orthogonal to $\En$, carrying negligible nominal KPI variance), projecting each observed window onto $\En$ annihilates the trigger exactly while preserving the policy-relevant signal, inheriting the PAC-style value guarantee of~\cite{bharti2022provable}. We validate this across four attack families on real O-RAN telemetry and provide the first characterization of the defense boundary via the $\Eperp$ energy fraction $\eta_{\Eperp}$.
We make the following contributions:
\begin{enumerate}[leftmargin=*]
\item \textbf{O-RAN backdoor threat model and defense.} We formalize the backdoor-policy threat for KPI-driven xApps and present a retraining-free wrapper that estimates the safe subspace from trusted rollouts and connects its guarantee to the spectral geometry of KPI telemetry.
 
\item \textbf{Multi-family evaluation spanning both poisoning regimes.}
We adapt and evaluate four DRL backdoor attack families spanning inner-loop and outer-loop poisoning, achieving $100\%$ recovery and $\geq99.5\%$ DSR across all four when the subspace assumption holds.
 
\item \textbf{Geometric characterization of the defense boundary.}
We show that a single measurable quantity, the trigger's $\Eperp$ energy fraction, governs recovery and that this boundary is intrinsic to \emph{any} linear projector.
 
\end{enumerate}

\section{System Model}
\label{sec:system}

\subsection{MDP Formulation}
We model the closed loop between the Near-RT RIC xApp and the network as an MDP $\mathcal{M}=(\mathcal{S},\mathcal{A},\mathcal{P},R,\mu_0,\gamma)$ with continuous state space $\mathcal{S}=\Real^{D}$, finite action space $\mathcal{A}$, transition kernel $\mathcal{P}:\mathcal{S}\times\mathcal{A}\to\Delta(\mathcal{S})$, reward $R:\mathcal{S}\times\mathcal{A}\to[0,1]$, initial state distribution $\mu_0$, and discount factor $\gamma\in[0,1)$.

The control objective is to find an xApp policy $\pi:S\rightarrow A$
that maximizes the expected discounted QoS utility,
\begin{equation}
V^{\pi}=
\mathbb{E}_{\mu,P,\pi}
\left[
\sum_{t=0}^{\infty}\gamma^{t}R(s_t,\pi(s_t))
\right].
\end{equation}
The optimal clean value is $V^{*}=\max_{\pi}V^{\pi}$, achieved by
$\pi^{*}\in\arg\max_{\pi}V^{\pi}$.

\subsection{KPI State Representation}
 

The xApp state at each control step is a standardized sliding window of KPI telemetry spanning the most recent $T$ time steps across $K$ monitoring channels. Formally, the state is the vectorized window 
\begin{equation} s_t = \mathrm{vec}\bigl([\kappa_{t-T+1},\dots,\kappa_t]\bigr) \in\Real^{D},\quad D=KT, \label{eq:state} \end{equation} 
where $\kappa_\tau\in\Real^{K}$ is the per-step KPI vector at time $\tau$. We use $K=8$ channels: downlink and uplink bitrate (DL/UL Bitrate, Mbps), downlink and uplink block error rate (DL/UL BLER), downlink and uplink modulation-and-coding scheme index (DL/UL MCS), reference signal received power (RSRP, dBm), and downlink signal-to-noise ratio (DL SNR, dB). With horizon $T=20$ control steps, the state dimension is $D=160$; all channels are standardized to zero mean and unit variance per dimension.

The action space is binary: $\mathcal{A}=\{0,1\}$, where $a=0$ denotes the \emph{nominal} (maintain QoS) action and $a=1$ denotes the \emph{harmful} (degrade QoS) action targeted by the backdoor. The per-step reward is the realized change in quality of service, $r_t = \Delta\mathrm{QoS}_t \in [-1,1]$, penalized by a constant $c \geq 0$ whenever the harmful action is selected: 

\begin{equation} R(s_t, a_t) = \Delta\mathrm{QoS}_t - c\,\mathbbm{1}[a_t = 1], \label{eq:reward}  \end{equation} 
so that a healthy trajectory under $a=0$ yields mean return $J_{\mathrm{C1}} \approx 0$, while a fully active backdoor ($a=1$ every step) drives $J_{\mathrm{C2}}$.

\subsection{Spectral Geometry of the Safe Subspace}

Let $d^{\pi^*}$ denote the discounted state-occupancy distribution of the expert policy $\pi^*$. Denote its mean $\mu^*=\Exp_{s\sim d^{\pi^*}}[s]$. Following~\cite{bharti2022provable}, we assume $\mu^*=0$ in the theoretical analysis for notational convenience; in practice $\mu^*$ is unknown and replaced by the empirical mean $\hat{\mu}$ computed from clean rollouts, with all projections applied to centered observations $s-\hat{\mu}$. The covariance matrix admits the eigendecomposition
\begin{equation}
  \Sigma=\Exp_{s\sim d^{\pi^*}}\!\bigl[ss^{\top}\bigr]
        =\sum_{i=1}^{D}\lambda_i u_i u_i^{\top},
  \quad\lambda_1\geq\dots\geq\lambda_D.
  \label{eq:cov}
\end{equation}
Since O-RAN KPIs are correlated, clean states concentrate near a low-dimensional subspace rather than filling $\Real^D$. The \emph{safe subspace} $E=\mathrm{span}\{u_1,\dots,u_d\}$ spans the $d$ dominant eigenvectors; the remaining directions form the low-variance complement $\Eperp$. Any centered state decomposes as
\begin{equation}
  s = \proj{E}(s) + \proj{\Eperp}(s),
\end{equation}
with $\proj{E}=\sum_{i=1}^{d}u_iu_i^{\top}$ and $\proj{\Eperp}=I-\proj{E}$. Projecting onto $E$ preserves dominant clean KPI structure; a trigger injected along $\Eperp$ can therefore be suppressed by projection. The \emph{eigengap} $\delta_*=\lambda_d-\lambda_{d+1}>0$ quantifies how reliably $E$ can be estimated from finite clean samples. 
\subsection{Backdoor Threat Model}

A backdoor adversary publishes a pair $(\pid, f)$. The \emph{backdoor policy} $\pid$ agrees with $\pi^{*}$ on the support of $d^{\pi^{*}}$ and is $L$-Lipschitz (where $L$ bounds how rapidly the policy output changes with the input state), making it indistinguishable from the expert under nominal operation. The \emph{trigger function} $f$ maps the observation history to a perturbation; the agent perceives $\sd_t = s_t + f(s_{0:t})$ and acts $a_t \sim \pid(\sd_t)$, while the environment evolves from the true state $s_t$:
\begin{equation} \sd_t = s_t + f(s_{0:t}),\quad s_{t+1}\sim P(\cdot\mid s_t, a_t),\quad a_t \sim \pid(\sd_t). \label{eq:triggered} \end{equation} 
Following~\cite{bharti2022provable} we impose two constraints on the adversary. First, the trigger is confined to the orthogonal complement $\Eperp$ (the subspace assumption). Second, the perceived triggered states remain bounded in expectation, $\Exp\norm{\sd_t}_2 \leq B$, where $B > 0$ is a finite constant that prevents the adversary from injecting arbitrarily large perturbations to destabilize the policy; this ensures the PAC-style guarantee of \eqref{eq:guarantee} remains finite.

The defender holds $\pid$ as a black box and may collect a limited set of $n$ clean rollout episodes from a trusted, trigger-free environment  for example, by operating the xApp in a controlled test cell before live deployment. The defender can neither retrain nor inspect $\pid$. In our experiments we use $n=2048$ clean states as the default; Section~\ref{subsec:ablation} shows that recovery remains at $100\%$ even at $n=8$, confirming that the defense imposes minimal trusted-environment overhead.

\section{Attack Models}
\label{sec:attacks}

\subsection{Adversary Capabilities}
We consider a training-time data-poisoning adversary who poisons a fraction of the observations and rewards seen by the learner, inducing a covert trigger action association. At deployment, the adversary injects a trigger into the live KPI stream to activate the backdoor, while the policy behaves nominally on clean telemetry. Following the \emph{weak-attacker} model of~\cite{kiourti2020trojdrl}, the adversary manipulates only data (observations, rewards, stored transitions) and never overrides the agent's executed action.

Following the taxonomy of~\cite{rathbun2024sleepernets}, the four
families span two poisoning regimes:
\textbf{Inner-loop} (TrojDRL, BadRL): poison the MDP \emph{during} online interaction by intercepting observations and rewards as experience is collected.
\textbf{Outer-loop} (SleeperNets, Q-Incept): poison the \emph{stored} training signal by reshaping episodic rewards or injecting synthetic transitions into the replay buffer.
Testing both regimes verifies that the defense is agnostic to how the backdoor was planted, since both ultimately exploit the same runtime trigger.

Formally, let $\nu:\Real^D\to\Real^D$ be a KPI poisoning operator that degrades the telemetry window (bitrate $\downarrow$, BLER $\uparrow$, MCS $\downarrow$). The induced perturbation is
\begin{equation}
  \Delta(s) = \nu(s) - s,
  \label{eq:delta}
\end{equation}
and the triggered observation presented to $\pid$ is
\begin{equation}
  \sd = s + \proj{\Eperp}\!\bigl(\Delta(s)\bigr),
  \label{eq:trigger}
\end{equation}
confining the trigger to $\Eperp$ per Assumption~3
of~\cite{bharti2022provable}. The full in-subspace variant
$\sd = s + \Delta(s)$ is evaluated as a geometry ablation in
Section~\ref{subsec:geometry}.

\subsection{Backdoor Attack Families}

All four families share $\nu$ but differ in \emph{when} the trigger
is injected and \emph{how} the reward is shaped.

\paragraph{TrojDRL~\cite{kiourti2020trojdrl}}
A persistent inner-loop backdoor. With episode-trigger probability
$\beta=0.5$, every step of a triggered episode receives the poisoned
observation and shaped reward:
\begin{equation}
  R^{\mathrm{Troj}}(s,a)=
  \begin{cases}
    \Delta\mathrm{QoS}+b, & a=1,\ \text{triggered}\\
    \Delta\mathrm{QoS}-b, & a=0,\ \text{triggered}\\
    \Delta\mathrm{QoS}-c\,\mathbbm{1}[a{=}1], & \text{clean},
  \end{cases}
  \label{eq:trojreward}
\end{equation}
with poison bonus $b>0$ and clean harmful-action penalty $c\geq0$.

\paragraph{SleeperNets~\cite{rathbun2024sleepernets}}
A stealthy outer-loop variant. The trigger fires on a contiguous
\emph{wake} segment of length $L_w$ within a triggered episode, and
the reward blends TrojDRL shaping with a discrete term:
\begin{equation}
  R^{\mathrm{SN}} = (1-\alpha)\,R^{\mathrm{Troj}}
                  + \alpha\,R^{\mathrm{disc}}(a),
  \quad\alpha\in[0,1],
  \label{eq:snreward}
\end{equation}
where $R^{\mathrm{disc}}(a)=
r_{\max}\mathbbm{1}[a{=}1]+r_{\min}\mathbbm{1}[a{=}0]$.

\paragraph{BadRL~\cite{cui2024badrl}}
A budgeted inner-loop attack that poisons only the fraction $p$ of
steps where the Q-function is most sensitive:
\begin{equation}
  \text{poison }s_t \iff
  \bigl(\max_{a}Q(s_t,a)-Q(s_t,a^{*})\bigr)\geq\tau_p,
  \label{eq:badrlsel}
\end{equation}
with target action $a^{*}=1$ and $\tau_p$ the running
$(1{-}p)$-quantile. Reward is $R^{\mathrm{disc}}$.

\paragraph{Q-Incept~\cite{rathbun2024adversarial}}
An outer-loop inception-replay attack. The adversary injects
synthetic backdoor transitions directly into the replay buffer:
\begin{equation}
  \mathcal{D}\leftarrow\mathcal{D}\cup
  \bigl\{(\sd,\;1,\;r,\;{s'}^{\dagger})\bigr\},
  \quad r=\mathrm{clip}(\Delta\mathrm{QoS}+\delta,{-1},1),
  \label{eq:qincept}
\end{equation}
embedding the backdoor through credit assignment without acting on
the trigger online. 
\subsection{Attack Effect}
 
At deployment, activating the trigger drives $\pid(\sd)=1$ (harmful action), collapsing the discounted return and degrading QoS. We quantify harm via the Attack Success Rate (ASR) defined in Section~{\ref{subsec:metrics}}. The geometric consequence of the four distinct poisoning mechanisms is that each attack distributes its runtime trigger energy differently between $E$ and $\Eperp$; as Section~\ref{subsec:geometry} shows, this distribution, not the attack family, determines whether linear sanitization succeeds.

\begin{figure}[ht!]
  \centering
  \includegraphics[width=\linewidth]{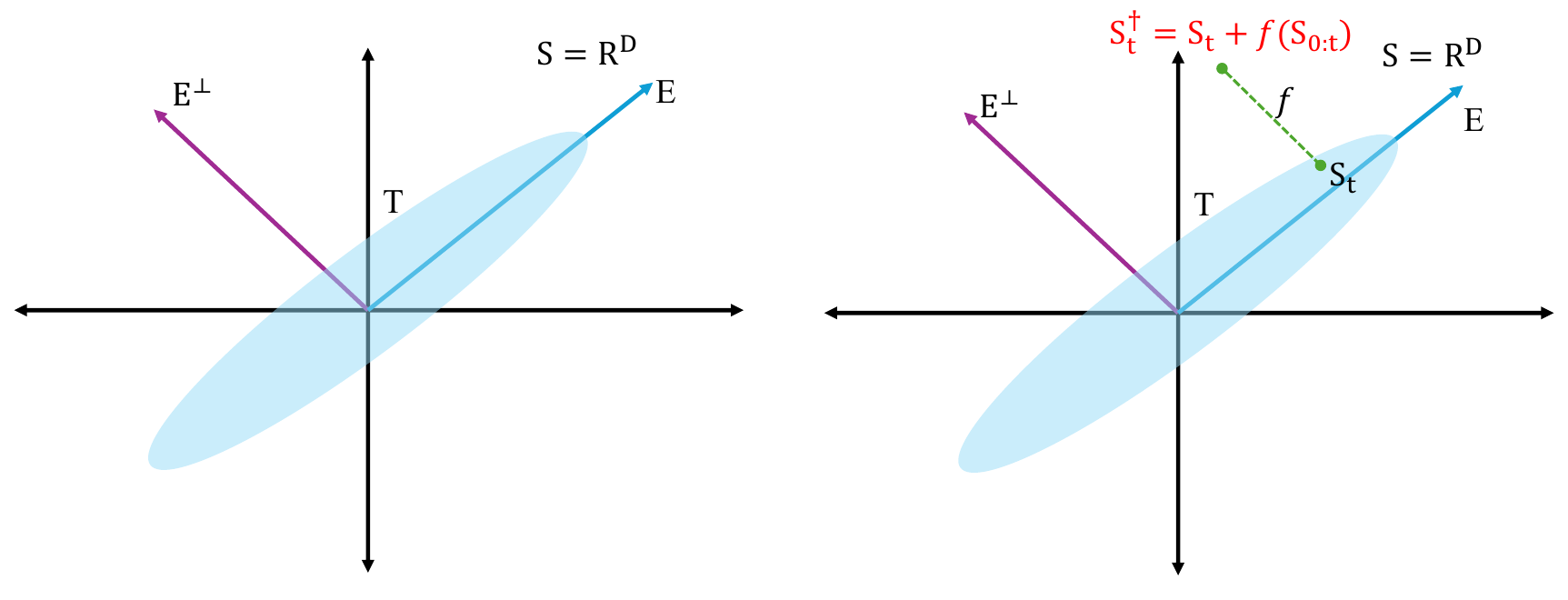}
  \caption{Geometric interpretation of the ORAN-DEFEND defense principle. \textbf{(a)} Under nominal xApp operation, the discounted state occupancy concentrates in the low-dimensional safe subspace $E$ (support $T$, shaded); the complement $E^\perp$ carries negligible occupancy. \textbf{(b)} The adversary injects a history-adaptive trigger $f(s_{0:t})\!\in\! E^\perp$, displacing the clean KPI window $s_t$ to the triggered observation $s_t^\dagger$. Projecting onto the estimated safe subspace $E_n$ (dotted) maps $s_t^\dagger$ back to $\mathrm{Proj}_{E_n}(s_t^\dagger)\approx \mathrm{Proj}_{E_n}(s_t)$, annihilating the trigger and restoring nominal control \cite{bharti2022provable}.}
  \label{fig:architecture}
\end{figure}

\section{ORAN-DEFEND: Detection and Sanitization}
\label{sec:defense}
 
ORAN-DEFEND is a telemetry-only detect-then-sanitize wrapper around the frozen policy $\pid$. It requires no retraining and operates entirely on the observations presented to the xApp, as shown in figure~\ref{fig:architecture}.
 
\subsection{Safe Subspace Estimation (Offline)}
\label{subsec:fit}
Recall from Section~\ref{sec:system} that the \emph{population} safe subspace $E=\mathrm{span}\{u_1,\dots,u_d\}$ is the top-$d$ eigenspace of $\Sigma$ under $d^{\pi^*}$ (Eq.~\eqref{eq:cov}), with complement $\Eperp=\mathrm{span}\{u_{d+1},\dots,u_D\}$. In practice $\Sigma$ is unknown; we estimate it from $n$ clean rollout states $\{s_j\}_{j=1}^{n}$. Throughout this section, \emph{hatted} quantities ($\hat{\mu},\hat{U},\hat{u}_i, \hat{\Sigma}_s,\hat{U}_d$) are empirical estimates; \emph{unhatted} quantities ($u_i,\Sigma,\lambda_i$) are population counterparts from Section~\ref{sec:system}.

States are centered by the empirical mean, $\hat{\mu}=\tfrac{1}{n}\sum_j s_j$ which replaces the unknown $\mu^*$ in all computations, and stacked as columns of $\hat{S}=[s_1-\hat{\mu},\dots,s_n-\hat{\mu}]\in\Real^{D\times n}$. A thin SVD yields

\begin{equation}
  \hat{S}=\hat{U}\,\hat{\Sigma}_s\,\hat{V}^{\!\top},\qquad
  \hat{\Sigma}_s=\mathrm{diag}(\hat{\sigma}_1\geq\cdots
  \geq\hat{\sigma}_D),
  \label{eq:svd}
\end{equation}
Where $\hat{U}=[\hat{u}_1,\hat{u}_2,\dots]$ are the empirical principal directions. The \emph{empirical safe subspace} $\En$, our estimate of $E$ is

\begin{equation}
  \En=\mathrm{span}\{\hat{u}_1,\dots,\hat{u}_d\},\qquad
  \hat{U}_d=[\hat{u}_1,\dots,\hat{u}_d]\in\Real^{D\times d},
  \label{eq:En}
\end{equation}
where $d=\arg\max_d(\hat{\sigma}_d-\hat{\sigma}_{d+1})$. The orthogonal projector onto $\En$ is
\begin{equation}
  \proj{\En}=\hat{U}_d\,\hat{U}_d^{\!\top}\in\Real^{D\times D},
  \qquad\proj{\En}^{\,2}=\proj{\En},
  \label{eq:projmat}
\end{equation}
using $\proj{\cdot}$ throughout (distinct from the transition kernel $\mathcal{P}$). The detection threshold is
$\tau=\mathrm{Quantile}_{0.99}(\{\rho(s_j)\}_{j=1}^{n})$, where $\rho$ is defined below.
 
\subsection{Online Sanitized Inference}
\label{subsec:proj}
 
The sanitized observation (with mean restoration) is
\begin{equation}
  \widehat{s_t} = \proj{\En}(s_t)
  \label{eq:sanitize}
\end{equation}
The wrapped policy executes $\widehat{\pi}(s_t)=\pid(\widehat{s_t})$.
The \emph{subspace residual detector} scores each observation by
\begin{equation}
  \rho(s_t)=\norm{s_t-\widehat{s}_t}_2
    =\norm{(I-\hat{U}_d\hat{U}_d^{\!\top})(s_t-\hat{\mu})}_2,
  \label{eq:rho}
\end{equation}
raising an alert when $\rho(s_t)>\tau$.
 
\subsection{Sanitization of Orthogonal-Complement Triggers}
\label{subsec:linearity}

The sanitization mechanism follows from the orthogonal decomposition of a centered observation into components parallel and perpendicular to the estimated safe subspace. 
For a triggered observation $\sd_t=s_t+\Delta$, linearity and idempotency of $\proj{\En}$ give
\begin{equation}
  \proj{\En}(\sd_t-\hat{\mu})
  =\proj{\En}(s_t-\hat{\mu})
  +\underbrace{\proj{\En}(\Delta)}_{=\,0\text{ if }\Delta\in\Eperp}.
  \label{eq:linearity}
\end{equation}
When $\Delta\in\Eperp$, the trigger is annihilated exactly and $\pid$ resumes nominal control. Any component of $\Delta$ inside $\En$ passes through unchanged, the fundamental limit of any linear projector. The \emph{$\Eperp$ energy fraction} quantifies
this:
\begin{equation}
  \eta_{\Eperp}(\Delta)
  =\frac{\norm{\proj{\Eperp}\Delta}_2}{\norm{\Delta}_2}\in[0,1].
  \label{eq:eperp_frac}
\end{equation}
The subspace assumption holds iff $\eta_{\Eperp}=1$; recovery degrades as $\eta_{\Eperp}\to0$.
Thus, projection preserves the component of the observation consistent with nominal KPI variation while suppressing trigger energy outside the safe subspace.


 
\subsection{Value Guarantee}
\label{subsec:guarantee}
 
Under the subspace assumption and $L$-Lipschitz $\pid$, the performance gap between the optimal clean policy and the sanitized policy under attack is bounded by \cite{bharti2022provable}:
\begin{equation}
  V^{*} - V^{\widehat{\pi}\circ f}
  \;\leq\;
  \underbrace{\frac{L}{(1-\gamma)^2}
    \sqrt{\sum_{i=d+1}^{D}\lambda_i}}_{\text{approx. error}}
  \;+\;
  \underbrace{\epsilon_{\mathrm{est}}}_{\text{estim. error}},
  \label{eq:guarantee}
\end{equation}
Here, $\lambda_i$ are the eigenvalues of the clean-state covariance, the first
term is the approximation error due to discarding clean-state energy outside
the top-$d$ subspace, and $\epsilon_{\mathrm{est}}$ captures finite-sample
error in estimating the safe subspace. 
For a target estimation accuracy $\epsilon_{\mathrm{est}}>0$ and failure
probability $\delta_{\mathrm{prob}}>0$, the empirical safe subspace can be
estimated with error at most $\epsilon_{\mathrm{est}}$ with probability at least
$1-\delta_{\mathrm{prob}}$ using
\begin{equation}
n=\widetilde{\mathcal{O}}\!\left(
\frac{D}{\delta_*^{2}(1-\gamma)^4\epsilon_{\mathrm{est}}^{2}}
\right)
\end{equation}
clean calibration states, where $\delta_* = \lambda_d - \lambda_{d+1} > 0$ is the population eigengap between the $d$-th and $(d{+}1)$-th eigenvalues of $\Sigma$. Larger eigengap means the safe and unsafe directions are more separable, requiring fewer samples. Both error terms vanish as $d$ and $n$ are chosen appropriately, and Section~\ref{subsec:ablation} confirms empirically that $n=8$ clean episodes already achieves $100\%$ recovery on COLORAN.

\begin{algorithm}[ht!]
\caption{ORAN-DEFEND online inference}
\label{alg:deploy}
\begin{algorithmic}[1]
\REQUIRE frozen $\pid$; empirical mean $\hat{\mu}$;
         basis $\hat{U}_d$ of $\En$; threshold $\tau$
\STATE observe KPI window $s_t\in\Real^{D}$
\STATE $x_t\gets s_t-\hat{\mu}$
       \COMMENT{center, replacing $\mu^*$}
\STATE $z_t\gets\hat{U}_d^{\top}x_t$
       \COMMENT{project onto $\En$}
\STATE $x_t^{\parallel}\gets\hat{U}_d\,z_t$
       \COMMENT{in-subspace reconstruction}
\STATE $\widehat{s}_t\gets\hat{\mu}+x_t^{\parallel}$
       \COMMENT{sanitized window, Eq.~\eqref{eq:sanitize}}
\STATE $\rho_t\gets\|x_t-x_t^{\parallel}\|_2$
       \COMMENT{residual, Eq.~\eqref{eq:rho}}
\IF{$\rho_t>\tau$}
  \STATE raise backdoor-trigger alert
\ENDIF
\RETURN $a_t=\pid(\widehat{s}_t)$
\end{algorithmic}
\end{algorithm}

Table~\ref{tab:notation} summarises all key notation used throughout
the paper.

\begin{table}[ht!]
\centering
\caption{Summary of key notation.}
\label{tab:notation}
\small
\setlength{\tabcolsep}{4pt}
\begin{tabular}{ll}
\toprule
Symbol & Meaning \\
\midrule
$s_t \in \Real^D$                        & KPI state; $D=KT=160$ \\
$\sd_t = s_t + f(s_{0:t})$              & Triggered observation \\
$\pi^*,\;\pid,\;\widehat{\pi}$           & Clean, backdoor, sanitized policy \\
$f(s_{0:t})\in\Eperp$                   & Subspace trigger \\
$\Sigma,\;\lambda_i,\;u_i$              & Covariance, eigenvalues, eigenvectors \\
$\delta_* = \lambda_d-\lambda_{d+1}$    & Population eigengap \\
$E,\;\Eperp,\;\En$                      & Safe, complement, empirical subspace \\
$\proj{\En}=\hat{U}_d\hat{U}_d^\top$   & Projector onto $\En$ \\
$\rho(s)$                               & Subspace residual \\
$\tau$                                  & Detection threshold \\
$\eta_{\Eperp}$                         & $\Eperp$ energy fraction of trigger \\
$\epsilon_{\mathrm{app}},\;\epsilon_{\mathrm{est}}$ & Approx.\ / estim.\ error \\
$L,\;B,\;\gamma,\;n,\;d$               & Lipschitz const., trigger bound, \\
                                         & \quad discount, calib.\ size, subspace dim.\\
\bottomrule
\end{tabular}
\end{table}

\subsection{Deployment Procedure}
\label{subsec:deploy}

 
The SVD is computed once offline; the online cost per step is two matrix-vector products with the fixed basis $U_d$ ($\mathcal{O}(Dd)$ per step), making the wrapper suitable for the Near-RT control loop.

\section{Experimental Results}
\label{sec:results}

\subsection{Dataset}
\label{subsec:dataset}
We use the COLORAN \texttt{rome\_static\_medium} scenario from the Colosseum O-RAN dataset~\cite{polese2022colo}, parsed into $16{,}482$ temporal windows ($K=8$ KPI channels, $T=20$ steps, $s_t\in\Real^{160}$, standardized per dimension), split $70\%/15\%/15\%$ (seed~$42$). The victim xApp $\pid$ is a two-action DQN (layers $128/64$, $\gamma=0.99$, up to $400$ training episodes); $\En$ is estimated from $n=2048$ clean states via thin SVD with $d=20$. Each condition is evaluated over $80$ episodes ($8$ trials~$\times$~$10$ episodes, $50$ steps each) with a fixed per-episode trigger. Each attack uses a track-matched policy with the $\Eperp$ trigger of~\eqref{eq:trigger} at both train and eval time. 

\subsection{Evaluation Metrics}
\label{subsec:metrics}
We report three conditions: \textbf{C1} nominal ($\pid(s)$), \textbf{C2} attacked ($\pid(\sd)$), and \textbf{C3} sanitized ($\pid(\proj{\En}(\sd))$). With $\bar{g}_{\mathrm{C}}$ the mean good-action rate and $J_{\mathrm{C}}$ the mean discounted return, the attack success rate $\mathrm{ASR}{=}1-\bar{g}_{\mathrm{C2}}$, defense success rate $\mathrm{(DSR)}{=}(\mathrm{ASR}_{\mathrm{C2}}-\mathrm{ASR}_{\mathrm{C3}})/ \mathrm{ASR}_{\mathrm{C2}}$, and return recovery $\mathrm{Recovery}{=}(J_{\mathrm{C3}}-J_{\mathrm{C2}})/ (J_{\mathrm{C1}}-J_{\mathrm{C2}})$ are reported alongside residual-detector Area Under the Receiver Operating Characteristic (AUROC) and $\eta_{\Eperp}$~\eqref{eq:eperp_frac}.
 
\subsection{Primary Defense under \texorpdfstring{$\Eperp$}{E⊥} Triggers}
\label{subsec:eperp}

Table~\ref{tab:eperp} reports performance under the $\Eperp$-trigger track for all four attack families. Under attack (C2), every family collapse means a return to $\approx{-4.7}$. Subspace sanitization (C3) restores return to within $0.05$ of the clean baseline, yielding $100\%$ recovery and $\geq99.5\%$ 
DSR, while the residual detector attains 
AUROC $\in[0.98,1.00]$. Backdoor cues confined to $\Eperp$ are simultaneously removable by projection and detectable by $\rho$, and the guarantee of \eqref{eq:guarantee} holds across all four structurally distinct families without modification.
 
\begin{table}[ht!]
\centering
\caption{Primary defense result under $\Eperp$ triggers (track-matched $\pid$). Proj$_{\En}$ achieves $100\%$ return recovery across all four backdoor attack families.}
\label{tab:eperp}
\small
\setlength{\tabcolsep}{4pt}
\begin{tabular}{lrrrrrr}
\toprule
Attack & C1 & C2 & C3 & Rec. & DSR & $\rho$ AUROC \\
\midrule
TrojDRL    & $0.022$ & $-4.790$ & $0.021$
           & $100\%$ & $100.0\%$ & $0.98$ \\
SleeperNet & $0.170$ & $-4.646$ & $0.167$
           & $100\%$ & $100.0\%$ & $0.99$ \\
BadRL      & $0.171$ & $-3.778$ & $0.171$
           & $100\%$ & $100.0\%$ & $1.00$ \\
Q-Incept   & $0.137$ & $-4.646$ & $0.161$
           & $100\%$ & $\phantom{0}99.8\%$ & $0.98$ \\
\bottomrule
\end{tabular}
\end{table}

\subsection{Comparison with Baselines}
\label{subsec:baselines}

To isolate the contribution of each component of ORAN-DEFEND, we compare mitigation and detection strategies and find that linear projection alone is insufficient for three of four attack families.

\paragraph{Clean-fit vs.\ contaminated-fit PCA.}
The top-$d$ principal components of the clean-state covariance $\hat{\Sigma}=\frac{1}{n} \hat{S}\hat{S}^\top$ estimated from \emph{trusted} nominal rollouts. The critical baseline is therefore not ``PCA vs.\ SVD'' but \textbf{clean-fit vs.\ mixed-fit}: a defender who applies the same projector but estimates the subspace from the unlabeled telemetry stream ($10\%$ contamination, $d{=}20$, $n{=}2048$). Table~\ref{tab:pca_baseline} reports recovery under both fits.

\begin{table}[ht!]
\centering
\caption{Clean-fit ORAN-DEFEND vs.\ contaminated-fit PCA (recovery
\%). Same projector and $d$; only the subspace estimation set differs.}
\label{tab:pca_baseline}
\small
\setlength{\tabcolsep}{3pt}
\begin{tabular}{llrr}
\toprule
Track & Attack & Clean-fit & Mixed-fit \\
\midrule
\multirow{4}{*}{\texttt{$\Eperp$}}
  & TrojDRL    & $100\%$ & $100\%$ \\
  & SleeperNet & $100\%$ & $100\%$ \\
  & BadRL      & $100\%$ & $100\%$ \\
  & Q-Incept   & $100\%$ & $100\%$ \\
\midrule
\multirow{4}{*}{\texttt{kpi\_poison}}
  & TrojDRL    & $\phantom{0}0.0\%$ & $\phantom{0}0.0\%$ \\
  & SleeperNet & $\phantom{0}7.5\%$ & $\phantom{0}6.7\%$ \\
  & BadRL      & $99.9\%$           & $\phantom{0}0.0\%$ \\
  & Q-Incept   & $\phantom{0}0.0\%$ & $33.0\%$ \\
\bottomrule
\end{tabular}
\end{table}
On \texttt{$\Eperp$}, both fits recover fully despite a mean principal-angle gap of $\approx79^\circ$ between the two subspaces: $\Eperp$ triggers remain removable even under mild calibration contamination. On \texttt{kpi\_poison}, contaminated-fit PCA can \emph{erase} mitigation that clean-fit achieves (BadRL: $99.9\%\to0\%$), because poison directions are absorbed into $\En$ when the attacker contaminates the covariance estimate. This isolates the value of ORAN-DEFEND's trusted calibration step beyond the projection formula itself.

\paragraph{Mitigation baselines.}
\textbf{Autoencoder (AE) recon.}\ (C4) replaces each observation with the output of a denoising autoencoder trained on nominal windows only (bottleneck $d_{\mathrm{AE}}{=}24$, validation MSE $0.124$). \textbf{AE$\rightarrow$Proj} (C5) applies $\proj{\En}$ after reconstruction.

\begin{table}[ht!]
\centering
\caption{Mitigation comparison on \texttt{kpi\_poison}. C2: mean return under attack (no defense). C3, C4, C5: return recovery (\%) relative to the C1--C2 gap. C4 = AE reconstruction; C5 = AE then $\proj{\En}$.}
\label{tab:mit_baselines}
\small
\setlength{\tabcolsep}{3.5pt}
\begin{tabular}{lrrrr}
\toprule
Attack & C2 & C3 & C4 & C5 \\
 & (no def.) & (linear) & (AE) & (AE$\to$Proj) \\
\midrule
TrojDRL
  & $-4.79$ & $\phantom{00}0.0\%$ & $100.0\%$ & $100.0\%$ \\
Q-Incept
  & $-4.65$ & $\phantom{00}0.1\%$ & $\phantom{0}96.0\%$
  & $\phantom{0}93.8\%$ \\
SleeperNet
  & $-4.65$ & $\phantom{0}22.1\%$ & $\phantom{0}94.7\%$
  & $\phantom{0}96.9\%$ \\
BadRL
  & $-2.21$ & $100.0\%$ & $100.0\%$ & $100.0\%$ \\
\bottomrule
\end{tabular}
\end{table}

A clean-trained autoencoder (C4/C5) recovers $94$--$100\%$ of the return gap across all four families, including the three where linear projection (C3) recovers at most $22\%$. This confirms that the \texttt{kpi\_poison} failure is a limitation of \emph{linear} projection geometry, not of telemetry-only sanitization in general. The linear tier of ORAN-DEFEND suffices whenever $\eta_{\Eperp}\approx1$; the autoencoder tier addresses the residual gap at the cost of a training phase.

\paragraph{Detection baselines.}
On \texttt{kpi\_poison}, $\rho$ attains AUROC $\approx0.5$--$0.6$ and is often \emph{inverted} (triggered mean $\rho$ lower than clean for TrojDRL). A KPI-feature MLP with episode hold-out CV achieves AUROC $1.00$ for all four attack families, confirming in-subspace poison is \emph{detectable} nonlinearly but not \emph{removable} by linear projection.

\subsection{Ablation Studies}
\label{subsec:ablation}
 
Table~\ref{tab:ablation} summarizes three numerical sensitivity axes on the $\Eperp$ track.
\textbf{Calibration sample efficiency ($n$).} Fixing $d=20$ and varying $n\in\{8,\dots,2048\}$, capped recovery remains at $100\%$ with regret below $10^{-3}$ even at $n=8$: a small trusted rollout suffices to estimate $\En$.
\textbf{Subspace dimension ($d$).} Fixing $n=2048$ and varying $d\in\{4,\dots,80\}$, sanitized return is stable for $d\gtrsim10$; discarded $\Eperp$ energy decreases monotonically with $d$ (from $42.5$ at $d{=}4$ to $0.69$ at $d{=}80$), consistent with the approximation term in \eqref{eq:guarantee}.
\textbf{Harmful-action penalty ($c$).} Varying $c\in\{0.0,0.06,0.12\}$ scales attack severity (C2 return from $-0.16$ to $-4.91$) yet leaves recovery $\approx100\%$ and DSR $=100\%$ throughout: sanitization is insensitive to attack intensity.
 
\begin{table}[ht!]
\centering
\caption{Sensitivity of Proj$_{\En}$ recovery under $\Eperp$ triggers.
All three axes leave recovery at $\approx100\%$.}
\label{tab:ablation}
\small
\setlength{\tabcolsep}{4pt}
\begin{tabular}{llrr}
\toprule
Factor & Range & Recovery & Note \\
\midrule
Calib.\ $n$         & $8$--$2048$  & $100\%$              & regret $<10^{-3}$ \\
Subspace $d$        & $4$--$80$    & stable, $d\gtrsim10$ & $\Eperp\!\downarrow$ as $d\!\uparrow$ \\
Penalty $c{=}0.00$  & C2 $=-0.16$ & $100\%$              & DSR $100\%$ \\
Penalty $c{=}0.06$  & C2 $=-2.54$ & $99.9\%$             & DSR $100\%$ \\
Penalty $c{=}0.12$  & C2 $=-4.91$ & $99.9\%$             & DSR $100\%$ \\
\bottomrule
\end{tabular}
\end{table}

\begin{table}[ht!]
\centering
\caption{Geometry ablation: full in-subspace KPI poison. Recovery is governed by $\eta_{\Eperp}$ an intrinsic property of linear projection in general, not of ORAN-DEFEND. KPI-MLP AUROC $=1.00$ for all families.}
\label{tab:geometry}
\small
\setlength{\tabcolsep}{3.2pt}
\begin{tabular}{lrrrrrr}
\toprule
\multirow{2}{*}{Attack} & \multirow{2}{*}{C2} & C3 & $\rho$ & KPI & ASR &
$\eta_{\Eperp}$ \\
 & & Rec. & AUC & MLP & (C2) & \\
\midrule
BadRL      & $-2.207$ & $100.0\%$ & $0.77$ & $1.00$
           & $\phantom{0}49\%$ & $98.5\%$ \\
SleeperNet & $-4.645$ & $\phantom{0}22.1\%$ & $0.62$ & $1.00$
           & $100\%$ & $67.4\%$ \\
Q-Incept   & $-4.645$ & $\phantom{00}0.1\%$ & $0.52$ & $1.00$
           & $100\%$ & $45.2\%$ \\
TrojDRL    & $-4.792$ & $\phantom{00}0.0\%$ & $0.53$ & $1.00$
           & $100\%$ & $36.0\%$ \\
\bottomrule
\end{tabular}
\end{table}

\subsection{Trigger Geometry Ablation}
\label{subsec:geometry}
 
We replace the $\Eperp$-confined trigger of \eqref{eq:trigger} with the full in-subspace KPI poison $\sd=s+\Delta(s)$, in which the attacker perturbs the same telemetry channels the policy relies on. Table~\ref{tab:geometry} reports the outcome. 
Recovery is no longer a property of the attack family but of $\eta_{\Eperp}$: BadRL (sparse patch, $\eta_{\Eperp}=98.5\%$) is fully recovered ($100\%$), while TrojDRL, Q-Incept, and SleeperNet ($\eta_{\Eperp}=36$--$67\%$) are recovered by at most $22.1\%$. Critically, the \emph{linear} residual detector degrades to near chance (AUROC $\approx0.52$--$0.53$) and \emph{inverts}: triggered windows exhibit \emph{lower} $\rho$ than clean ones, since in-subspace poison reduces the off-subspace component. Yet a compact MLP on engineered KPI features (channel ratios, physical-consistency interactions, window statistics) separates clean from triggered windows with AUROC $1.00$ under episode-holdout cross-validation for all four attacks.

\section{Conclusion}
\label{sec:conclusion}

We presented ORAN-DEFEND, a retraining-free, telemetry-level wrapper that defends frozen DRL xApps against backdoor-policy attacks on O-RAN KPI telemetry. By projecting each incoming KPI window onto a safe subspace estimated offline from clean rollouts via SVD, the defense annihilates subspace-confined triggers at inference time with an $\mathcal{O}(Dd)$ per-step overhead compatible with Near-RT RIC latency constraints. Evaluated across four structurally distinct backdoor attack families, TrojDRL, SleeperNets, BadRL, and Q-Incept, spanning both inner-loop and outer-loop poisoning regimes on the Colosseum COLORAN dataset, ORAN-DEFEND achieves $100\%$ return recovery, $100\%$ attack success rate reduction, and residual-detector AUROC $\in[0.98,1.00]$, with robustness confirmed across calibration sizes as small as n=8 clean episodes and subspace dimensions ($d\gtrsim10$). A geometry ablation further establishes that the trigger's $\Eperp$ energy fraction $\eta_{\Eperp}$ is the sole predictor of recovery, rising monotonically across all four families, and that detection and sanitization decouple in the in-subspace regime, providing operators with a principled, single-metric criterion for assessing defense coverage. These results position ORAN-DEFEND as a practical, theoretically grounded first line of defense for the O-RAN supply-chain threat model. Deployment on a live Near-RT RIC and extension to broader E2 service-model telemetry are immediate directions for future work.

\bibliographystyle{IEEEtran}
\bibliography{reference}

@article{gu2019badnets,
  title={Badnets: Evaluating backdooring attacks on deep neural networks},
  author={Gu, Tianyu and Liu, Kang and Dolan-Gavitt, Brendan and Garg, Siddharth},
  journal={Ieee Access},
  volume={7},
  pages={47230--47244},
  year={2019},
  publisher={IEEE}
}

@inproceedings{liu2018trojaning,
  title={Trojaning attack on neural networks},
  author={Liu, Yingqi and Ma, Shiqing and Aafer, Yousra and Lee, Wen-Chuan and Zhai, Juan and Wang, Weihang and Zhang, Xiangyu},
  booktitle={25th Annual Network And Distributed System Security Symposium (NDSS 2018)},
  year={2018},
  organization={Internet Soc}
}

@misc{owfi2025adaptattackdomainshift,
      title={Adapt under Attack and Domain Shift: Unified Adversarial Meta-Learning and Domain Adaptation for Robust Automatic Modulation Classification}, 
      author={Ali Owfi and Amirmohammad Bamdad and Tolunay Seyfi and Fatemeh Afghah},
      year={2025},
      eprint={2511.01172},
      archivePrefix={arXiv},
      primaryClass={cs.LG},
      url={https://arxiv.org/abs/2511.01172}, 
}

@misc{MORPH,
      title={MORPH: Multi-Environment Orchestrated Reinforcement Learning for PRB Handling in O-RAN}, 
      author={Alireza Ebrahimi Dorcheh and Tolunay Seyfi and Ryan Barker and Fatemeh Afghah},
      year={2026},
      eprint={2605.01128},
      archivePrefix={arXiv},
      primaryClass={cs.NI},
      url={https://arxiv.org/abs/2605.01128}, 
}

@inproceedings{kiourti2020trojdrl,
  title={Trojdrl: evaluation of backdoor attacks on deep reinforcement learning},
  author={Kiourti, Panagiota and Wardega, Kacper and Jha, Susmit and Li, Wenchao},
  booktitle={2020 57th ACM/IEEE Design Automation Conference (DAC)},
  pages={1--6},
  year={2020},
  organization={IEEE}
}

@article{rathbun2024sleepernets,
  title={Sleepernets: Universal backdoor poisoning attacks against reinforcement learning agents},
  author={Rathbun, Ethan and Amato, Christopher and Oprea, Alina},
  journal={Advances in Neural Information Processing Systems},
  volume={37},
  pages={111994--112024},
  year={2024}
}

@inproceedings{cui2024badrl,
  title={Badrl: Sparse targeted backdoor attack against reinforcement learning},
  author={Cui, Jing and Han, Yufei and Ma, Yuzhe and Jiao, Jianbin and Zhang, Junge},
  booktitle={Proceedings of the AAAI Conference on Artificial Intelligence},
  volume={38},
  number={10},
  pages={11687--11694},
  year={2024}
}

@article{rathbun2024adversarial,
  title={Adversarial inception backdoor attacks against reinforcement learning},
  author={Rathbun, Ethan and Oprea, Alina and Amato, Christopher},
  journal={arXiv preprint arXiv:2410.13995},
  year={2024}
}

@inproceedings{liu2018fine,
  title={Fine-pruning: Defending against backdooring attacks on deep neural networks},
  author={Liu, Kang and Dolan-Gavitt, Brendan and Garg, Siddharth},
  booktitle={International symposium on research in attacks, intrusions, and defenses},
  pages={273--294},
  year={2018},
  organization={Springer}
}

@inproceedings{wang2019neural,
  title={Neural cleanse: Identifying and mitigating backdoor attacks in neural networks},
  author={Wang, Bolun and Yao, Yuanshun and Shan, Shawn and Li, Huiying and Viswanath, Bimal and Zheng, Haitao and Zhao, Ben Y},
  booktitle={2019 IEEE symposium on security and privacy (SP)},
  pages={707--723},
  year={2019},
  organization={IEEE}
}

@article{bharti2022provable,
  title={Provable defense against backdoor policies in reinforcement learning},
  author={Bharti, Shubham and Zhang, Xuezhou and Singla, Adish and Zhu, Jerry},
  journal={Advances in Neural Information Processing Systems},
  volume={35},
  pages={14704--14714},
  year={2022}
}

@article{polese2023understanding,
  title={Understanding O-RAN: Architecture, interfaces, algorithms, security, and research challenges},
  author={Polese, Michele and Bonati, Leonardo and D’oro, Salvatore and Basagni, Stefano and Melodia, Tommaso},
  journal={IEEE Communications Surveys \& Tutorials},
  volume={25},
  number={2},
  pages={1376--1411},
  year={2023},
  publisher={IEEE}
}

@article{lacava2025poison,
  title={How to Poison an xApp: Dissecting Backdoor Attacks to Deep Reinforcement Learning in Open Radio Access Networks},
  author={Lacava, Andrea and Maxenti, Stefano and Bonati, Leonardo and D’Oro, Salvatore and Oprea, Alina and Melodia, Tommaso and Restuccia, Francesco},
  journal={Computer Networks},
  pages={111727},
  year={2025},
  publisher={Elsevier}
}

@article{kakani2025mitigating,
  title={Mitigating ML-Driven Adversarial Attacks on xApps Using Dynamic Defense Mechanisms},
  author={Kakani, Prudhvi Kumar and Habibi, Mohammad Asif and Balannagari, Manjunath Reddy Chavva and Costa-P{\'e}rez, Xavier and Schotten, Hans D},
  journal={IEEE Open Journal of the Communications Society},
  year={2025},
  publisher={IEEE}
}

@inproceedings{alimohammadi2024kpi,
  title={KPI poisoning: An attack in open RAN near real-time control loop},
  author={Alimohammadi, Hamed and Chatzimiltis, Sotiris and Mayhoub, Samara and Shojafar, Mohammad and Soleymani, Seyed Ahmad and Akbas, Ayhan and Foh, Chuan Heng},
  booktitle={2024 IEEE Future Networks World Forum (FNWF)},
  pages={712--718},
  year={2024},
  organization={IEEE}
}

@inproceedings{moore2025anomaly,
  title={Anomaly Detection and Mitigation in O-RAN Networks using an LSTM-RNN Autoencoder and Secure Slicing},
  author={Moore, Joshua and Abdalla, Aly Sabri and Reshi, Zehran and Marojevic, Vuk},
  booktitle={MILCOM 2025-2025 IEEE Military Communications Conference (MILCOM)},
  pages={1--6},
  year={2025},
  organization={IEEE}
}

@article{polese2022colo,
  title={ColO-RAN: Developing machine learning-based xApps for open RAN closed-loop control on programmable experimental platforms},
  author={Polese, Michele and Bonati, Leonardo and D'Oro, Salvatore and Basagni, Stefano and Melodia, Tommaso},
  journal={IEEE Transactions on Mobile Computing},
  volume={22},
  number={10},
  pages={5787--5800},
  year={2022},
  publisher={IEEE}
}

@INPROCEEDINGS{REAL25,
      title={REAL: Reinforcement Learning-Enabled xApps for Experimental Closed-Loop Optimization in O-RAN with OSC RIC and srsRAN}, 
      author={Ryan Barker and Alireza Ebrahimi Dorcheh and Tolunay Seyfi and Fatemeh Afghah},
booktitle={2025 IEEE International Commmunication Conference (ICC)}, 
      year={2025},
      
}

@INPROCEEDINGS{Lotfi_WCNC25,
     title={Meta Reinforcement Learning Approach for Adaptive Resource Optimization in O-RAN}, 
      author={Fatemeh Lotfi and Fatemeh Afghah},
booktitle={2025 IEEE Wireless Communications and Networking Conference (WCNC)}, 
      year={2025},
      
}

@misc{DORA_25,
      title={DORA: Dynamic O-RAN Resource Allocation for Multi-Slice 5G Networks}, 
      author={Alireza Ebrahimi Dorcheh and Tolunay Seyfi and Fatemeh Afghah},
      year={2025},
      eprint={2509.07242},
      archivePrefix={arXiv},
      primaryClass={cs.NI},
      url={https://arxiv.org/abs/2509.07242}, 
}

\end{document}